# The radiation problem from a vertical short dipole antenna above flat and lossy ground : novel formulation in the spectral domain with closed – form analytical solution in the high frequency regime

Ch. Christakis[1], K. Ioannidi[1], S.Sautbekov[2], P. Frangos[1] and S.K. Atanov[2]
[1]School of Electrical and Computer Engineering,
National Technical University of Athens,
9, Iroon Polytechniou Str., 157 73 Zografou, Athens, Greece
[2]Eurasian National University, 5, Munaitpassov Str., Astana, Kazakhstan
pfrangos@central.ntua.gr

*Abstract*—In this paper we consider the problem of radiation from a vertical short (Hertzian) dipole above flat lossy ground, which represents the well – known in the literature 'Sommerfeld radiation problem'. The problem is formulated in a novel spectral domain approach, and by inverse three – dimensional Fourier transformation the expressions for the received electric and magnetic (EM) field in the physical space are derived as one – dimensional integrals over the radial component of wavevector, in cylindrical coordinates. Subsequent use of the Stationary Phase Method (SPM) in the high frequency regime yields closed – form analytical solutions for the received EM field vectors, which coincide with the corresponding reflected EM field originating from the image point. In this way, we conclude that the so – called in the literature 'space wave' (line of sight plus reflected EM field) represents the total solution of the Sommerfeld problem in the high frequency regime, in which case the surface wave can be ignored. Finally, numerical results in the high frequency regime are presented in this paper, in comparison with corresponding numerical results based on Norton's solution of the problem (space and surface waves).

## I. Introduction

The so - called 'Sommerfeld radiation problem' is a well – known problem in the area of propagation of electromagnetic (EM) waves above flat lossy ground for obvious important applications in the area of wireless telecommunications [1,2]. The classical Sommerfeld solution to this problem is provided in the physical space by using the so- called 'Hertz potentials' and it does not end – up with closed form analytical solutions. K. A. Norton [3] concentrated in subsequent years more in the engineering application of the above problem with obvious application to wireless telecommunications, and he provided approximate solutions to the above problem, which are represented by rather long algebraic expressions for engineering use, in which the so – called 'attenuation coefficient' for the propagating surface wave plays an important role.

In this paper the authors take advantage of previous research work of them for the EM radiation problem in free space [4] by using the spectral domain approach.

Furthermore, in Ref. [5] the authors provided the fundamental formulation for the problem considered here, that is the solution in spectral domain for the radiation from a dipole moment at a specific angular frequency (ω) in isotropic media with a flat infinite interface. At that paper, the authors end – up with integral representations for the received electric and magnetic fields above or below the interface [Line of Site (LOS) plus scattered fields – transmitted fields, respectively], where the integration takes place over the radial spectral coordinate $k_\rho$, in cylindrical coordinates. Then, in the present paper the authors concentrate to the solution of the classical 'Sommerfeld radiation problem' described above *in the high frequency regime*, where the radiation from a vertical dipole moment at angular frequency ω takes place above flat lossy ground [this is equivalent to the radiation of a vertical small (Hertzian) dipole antenna above the flat lossy ground]. By using the *Stationary Phase Method* (SPM method [6], i.e. *high frequency approximation*) integration over the radial spectral coordinate $k_\rho$ is performed and closed – form analytical solutions for the received electric and magnetic (EM) fields are derived. These expressions coincide with the expressions corresponding to the EM field reflected from ground and originating from the image point, thus showing that this represents the total solution to the classical Sommerfeld radiation problem in the high frequency regime, where the surface wave can be ignored. Furthermore, the above derivations prove the accuracy of the expressions derived in this paper for the received EM field in spectral domain in all aspects. Finally, numerical results in the high frequency regime are presented in this paper, in comparison with corresponding numerical results based on Norton's solution of the problem (space and surface waves [3]). Full mathematical derivations of the expressions derived in this paper in the spectral domain, or by using, subsequently, the Stationary Phase Method (SPM) are provided by the authors in Ref. [7].

## II. GEOMETRY OF THE PROBLEM

The geometry of the problem is given in Fig. 1. Here a Hertzian (small) dipole with dipole moment p directed parallel to positive x – axis, at altitude $x_0$ above the infinite, flat and lossy ground, radiates time – harmonic electromagnetic (EM) waves at angular frequency ω=2πf [exp(-iωt) time dependence is assumed in this paper]. Here the relative complex permittivity of the ground (medium 2) is $\varepsilon'_r = \varepsilon'/\varepsilon_0 = \varepsilon_r + ix$, where $x = \sigma/\omega\varepsilon_0 = 18 \times 10^9 \, \sigma/f$, σ being the ground conductivity, f the frequency of radiation and $\varepsilon_0 = 8.854 \times 10^{-12}$ F/m is the absolute permittivity in vacuum or air. Then the wavenumbers of propagation of EM waves in air and lossy ground, respectively, are given by the following equations:

$$k_{01} = \omega/c_1 = \omega\sqrt{\varepsilon_1\mu_1} = \omega\sqrt{\varepsilon_0\mu_0} \tag{1}$$

$$k_{02} = \omega/c_2 = \omega\sqrt{\varepsilon_2\mu_2} = k_{01}\sqrt{\varepsilon_r + ix} \tag{2}$$

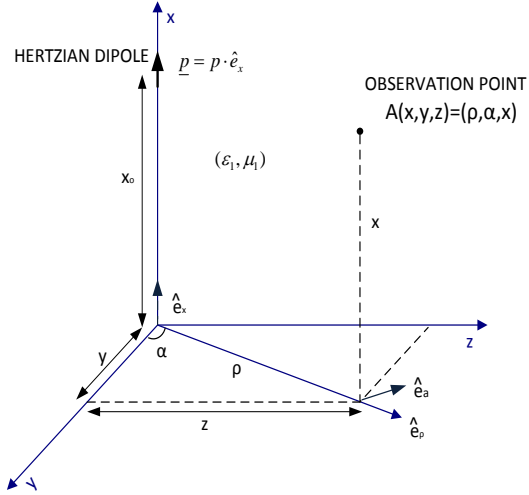

Fig. 1. Geometry of the problem

## III. EXPRESSIONS FOR THE RECEIVED ELECTROMAGNETIC (EM) FIELD OBTAINED THROUGH THE FORMULATION IN THE SPECTRAL DOMAIN

Following [5],[7],[8] and [9], we find the following expressions for the received EM fields [in the higher half - space, x>0, i.e. Line of Sight (LOS) field and field scattered from the ground, for receiver positions above it], as given below :

$$\underline{H}(\underline{r}) = \underline{H}^{LOS} - \frac{i\omega p\,\hat{e}_\alpha}{8\pi}\int_{-\infty}^{\infty}\frac{\varepsilon_{r2}\kappa_1 - \varepsilon_{r1}\kappa_2}{\kappa_1(\varepsilon_{r2}\kappa_1 + \varepsilon_{r1}\kappa_2)}k_\rho^2 \cdot H_0^{(1)}(k_\rho\rho)e^{i\kappa_1(x_0+x)}dk_\rho \tag{3}$$

$$\underline{E}(\underline{r}) = \underline{E}^{LOS}(\underline{r}) - \frac{ip}{8\pi\varepsilon_{r1}\varepsilon_0}\hat{e}_\rho\int_{-\infty}^{\infty}k_\rho^2\frac{\varepsilon_{r2}\kappa_1 - \varepsilon_{r1}\kappa_2}{(\varepsilon_{r2}\kappa_1 + \varepsilon_{r1}\kappa_2)}e^{i\kappa_1(x+x_0)}H_0^{(1)}(k_\rho\rho)dk_\rho +$$

$$+ \frac{ip}{8\pi\varepsilon_{r1}\varepsilon_0}\hat{e}_x\int_{-\infty}^{\infty}k_\rho^3\frac{\varepsilon_{r2}\kappa_1 - \varepsilon_{r1}\kappa_2}{\kappa_1(\varepsilon_{r2}\kappa_1 + \varepsilon_{r1}\kappa_2)}e^{i\kappa_1(x+x_0)}\cdot H_0^{(1)}(k_\rho\rho)dk_\rho \tag{4}$$

Similarly, the expressions for the received EM field in the lower half – space (x<0), i.e. EM field transmitted below the ground, are given by the following expressions (see [5] and [7] for mathematical details) :

$$\underline{H}^T(\underline{r}) = -\frac{i\omega p}{4\pi}\hat{e}_\alpha\int_{-\infty}^{\infty}k_\rho^2\frac{\varepsilon_{r2}}{\varepsilon_{r2}\kappa_1 + \varepsilon_{r1}\kappa_2}e^{i(\kappa_1 x_0 - \kappa_2 x)}\cdot H_0^{(1)}(k_\rho\rho)dk_\rho \tag{5}$$

$$\underline{E}^T(\underline{r}) = -\frac{ip}{4\pi\varepsilon_0}\int_{-\infty}^{\infty}(k_\rho\hat{e}_x - \kappa_2\hat{e}_\rho)\frac{k_\rho^2}{\varepsilon_{r2}\kappa_1 + \varepsilon_{r1}\kappa_2}\cdot e^{i(\kappa_1 x_0 - \kappa_2 x)}H_0^{(1)}(k_\rho\rho)dk_\rho \tag{6}$$

In eqs. (3) – (6) above $H_0^{(1)}$ is the Hankel function of first kind and zero order and

$$\kappa_1 = \sqrt{k_{01}^2 - k_\rho^2} \tag{7}$$

$$\kappa_2 = \sqrt{k_{02}^2 - k_\rho^2} \tag{8}$$

## IV. ANALYTICAL CLOSED- FORM EXPRESSIONS FOR THE SCATTERED EM FIELDS OBTAINED THROUGH THE APPLICATION OF THE STATIONARY PHASE METHOD (SPM)

Furthermore, by using the *Stationary Phase Method (SPM)* [which is a well – known method for calculating integrals *in the high – frequency regime* [6] ], we end - up with the following expressions for the EM field in the higher half – space ( x>0) [7]:

$$\underline{E}_{x>0} = \underline{E}^{LOS} - \frac{ip}{8\pi\varepsilon_o\varepsilon_{r1}}I_1\cdot\hat{e}_\rho - \frac{ip}{8\pi\varepsilon_o\varepsilon_{r1}}I_2\cdot\hat{e}_x \tag{9}$$

$$\underline{H}_{x>0} = \underline{H}^{LOS} - \frac{i\omega p}{8\pi}I_3\cdot\hat{e}_\alpha \tag{10}$$

where:

$$I_1 = \frac{i2}{k_{01}\rho^{1/2}}\frac{1}{(x+x_0)^{1/2}}\kappa_{1s}^{3/2}k_{\rho s}^{3/2}\cdot$$

$$\cdot\frac{\varepsilon_2\kappa_{1s} - \varepsilon_1\kappa_{2s}}{\varepsilon_2\kappa_{1s} + \varepsilon_1\kappa_{2s}}e^{ik_{\rho s}\rho}e^{i\kappa_{1s}(x+x_0)} \tag{11}$$

$$I_2 = \frac{i2}{k_{01}\rho^{1/2}}\frac{1}{(x+x_0)^{1/2}}\kappa_{1s}^{1/2}k_{\rho s}^{5/2}\cdot$$

$$\cdot\frac{\varepsilon_2\kappa_{1s} - \varepsilon_1\kappa_{2s}}{\varepsilon_2\kappa_{1s} + \varepsilon_1\kappa_{2s}}e^{ik_{\rho s}\rho}e^{i\kappa_{1s}(x+x_0)} \tag{12}$$

$$I_3 = \frac{i2}{k_{01}\rho^{1/2}}\frac{1}{(x+x_0)^{1/2}}\kappa_{1s}^{1/2}k_{\rho s}^{3/2}\cdot$$

$$\cdot\frac{\varepsilon_2\kappa_{1s} - \varepsilon_1\kappa_{2s}}{\varepsilon_2\kappa_{1s} + \varepsilon_1\kappa_{2s}}e^{ik_{\rho s}\rho}e^{i\kappa_{1s}(x+x_0)} \tag{13}$$

and

$$k_{\rho s} = \frac{k_{01}\rho}{\left[(x+x_0)^2 + \rho^2\right]^{1/2}} =$$

$$= k_{01}\frac{1}{\left[1+\left(\frac{x+x_0}{\rho}\right)^2\right]^{1/2}} = k_{01}\cos\phi \quad (14)$$

is the (unique) stationary point [7], where $\varphi$ is the angle defined by the image point of the radiating dipole, the observation point and the horizontal line drawn from the above mentioned image point, and $\cos\varphi$ is given by eq. (14) above (i.e. angle $\varphi$ is the well – known in the literature '*grazing angle*' [10]), as shown in Fig. 2, below.

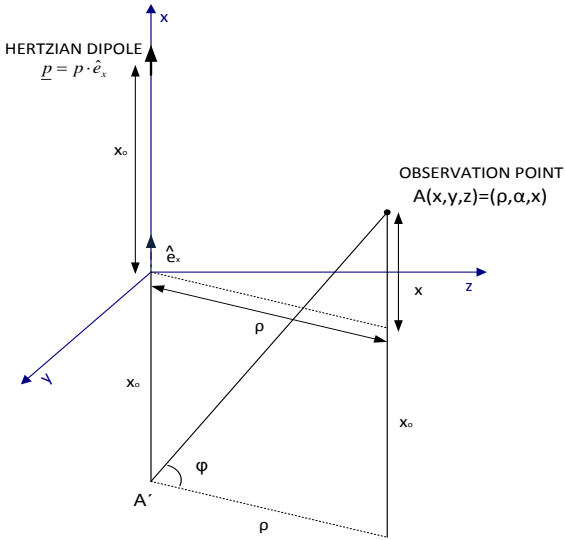

Fig. 2. Geometry of the radiation problem considered in this paper, where the image A' of the radiating Hertzian dipole is also shown.

Furthermore, regarding eqs. (11) - (13) above, the following quantities have been introduced :

$$\kappa_{1s} = \sqrt{k_{01}^2 - k_{\rho s}^2} = k_{01}\sin\phi \quad (15)$$

where angle $\varphi$ has been defined just above, and

$$\kappa_{2s} = \sqrt{k_{02}^2 - k_{\rho s}^2} \quad (16)$$

Finishing up the algebraic calculations in eqs. (11) – (13) above, we finally end – up with the following expressions for the received scattered electric and magnetic field vectors in the region x>0 :

$$\underline{E}_{x>0}^{sc} = \frac{p}{4\pi\varepsilon_o\varepsilon_{r1}}\frac{1}{\rho^{1/2}}\frac{1}{(x+x_0)^{1/2}}\frac{\kappa_{1s}^{1/2}k_{\rho s}^{3/2}}{k_{01}}\cdot\frac{\varepsilon_2\kappa_{1s}-\varepsilon_1\kappa_{2s}}{\varepsilon_2\kappa_{1s}+\varepsilon_1\kappa_{2s}}e^{ik_{\rho s}\rho}e^{i\kappa_{1s}(x+x_0)}\cdot(\kappa_{1s}\hat{e}_\rho + k_{\rho s}\hat{e}_x) =$$

$$= \frac{pk_{01}}{4\pi\varepsilon_o\varepsilon_{r1}}\frac{(\sin\phi)^{1/2}(\cos\phi)^{3/2}}{\rho^{1/2}(x+x_0)^{1/2}}\cdot\frac{\varepsilon_2\kappa_{1s}-\varepsilon_1\kappa_{2s}}{\varepsilon_2\kappa_{1s}+\varepsilon_1\kappa_{2s}}e^{ik_{\rho s}\rho}e^{i\kappa_{1s}(x+x_0)}\cdot(\kappa_{1s}\hat{e}_\rho + k_{\rho s}\hat{e}_x) =$$

$$= \frac{pk_{01}\cos\phi}{4\pi\varepsilon_o\varepsilon_{r1}(A'A)}\cdot\frac{\varepsilon_2\kappa_{1s}-\varepsilon_1\kappa_{2s}}{\varepsilon_2\kappa_{1s}+\varepsilon_1\kappa_{2s}}e^{ik_{\rho s}\rho}e^{i\kappa_{1s}(x+x_0)}\cdot(\kappa_{1s}\hat{e}_\rho + k_{\rho s}\hat{e}_x)$$

(17)

$$\underline{H}_{x>0}^{sc} = \frac{\omega p}{4\pi}\frac{1}{\rho^{1/2}}\frac{1}{(x+x_0)^{1/2}}\frac{\kappa_{1s}^{1/2}k_{\rho s}^{3/2}}{k_{01}}\cdot\frac{\varepsilon_2\kappa_{1s}-\varepsilon_1\kappa_{2s}}{\varepsilon_2\kappa_{1s}+\varepsilon_1\kappa_{2s}}e^{ik_{\rho s}\rho}e^{i\kappa_{1s}(x+x_0)}\hat{e}_\alpha =$$

$$= \frac{\omega k_{01}p}{4\pi}\frac{(\sin\phi)^{1/2}(\cos\phi)^{3/2}}{\rho^{1/2}(x+x_0)^{1/2}}\frac{\varepsilon_2\kappa_{1s}-\varepsilon_1\kappa_{2s}}{\varepsilon_2\kappa_{1s}+\varepsilon_1\kappa_{2s}}e^{ik_{\rho s}\rho}e^{i\kappa_{1s}(x+x_0)}\hat{e}_\alpha =$$

$$= \frac{\omega k_{01}p\cdot\cos\phi}{4\pi(A'A)}\frac{\varepsilon_2\kappa_{1s}-\varepsilon_1\kappa_{2s}}{\varepsilon_2\kappa_{1s}+\varepsilon_1\kappa_{2s}}e^{ik_{\rho s}\rho}e^{i\kappa_{1s}(x+x_0)}\hat{e}_\alpha$$

(18)

It can be easily derived from eqs. (17) and (18) above that :

$$\frac{\left|\underline{E}_{tot}^{sc}\right|_{x>0}}{\left|\underline{H}^{sc}\right|_{x>0}} = \zeta = \sqrt{\frac{\mu_0}{\varepsilon_0}} \quad (19)$$

where $\zeta$ is the free space impedance.

The above expressions (17) and (18) are the classical expressions for the EM field reflected from the lossy ground and originating from the image point, as shown in Fig. 2. Then, by using the newly derived in this paper expressions for the received EM field in spectral domain, eqs. (3) - (6) above, and by applying the SPM method, i.e. *in the high frequency regime*, the classical 'space wave' in region x>0 [10] was derived. This result has the following two interesting consequences :

1. The 'space wave' [10], which corresponds to the coherent summation (interference) of the reflected fields, eqs. (17) and (18) above, and the Line – of – Sight (LOS) fields (the latter not included in these equations) *is the solution to the Sommerfeld radiation problem in the high frequency regime*, where the so – called 'surface wave' can be ignored [3], [10].

2. The validity of expressions (3) and (4) above for the scattered EM field above the lossy and flat ground, derived in a novel way in this paper, has been confirmed in all aspects in the high frequency regime. Then it appears that expressions (3) – (6) above represent a very convenient starting point for further research with respect to the calculation of the received EM field for any frequency of the radiating dipole (i.e. including also low frequency effects), either above or below the ground, in an exact analytical way using the 'residue theorem' [11], or in a numerical way (i.e. through numerical integration techniques).

V. NUMERICAL RESULTS IN THE HIGH FREQUENCY REGIME – COMPARISON WITH NORTON'S RESULTS FOR SPACE AND SURFACE WAVES

In this Section indicative numerical results are provided for the electric field (magnitude) at the receiver point as a function of the horizontal distance (ρ) between transmitting Hertzian dipole and receiver position. These numerical results include the electric field scattered from the ground, magnitude of eq. (17), above, the Line – of – Sight (LOS) field, the so – called 'space wave' [which is just the coherent summation (interference) of the two fields mentioned above] and, finally, the so – called 'surface wave', according to Norton [3,10]. Furthermore, these numerical results are provided for frequency of radiating dipole f=80 MHz (Fig. 3) or f=30 MHz (Fig. 4). Note that at the higher frequency of 80 MHz (Fig. 3) the surface wave, according to Norton's formulation [3,10] is rather negligible, as compared to the 'space wave', while it becomes rather more important at the lower frequency of 30 MHz (Fig. 4). Our proposed SPM method of Section IV (which is inherently a 'high frequency method') ignores this surface wave contribution in the high frequency regime.

Moreover, note that the problem parameters in Figs. 3 and 4 are selected as following : height of transmitting dipole $x_0$=60m, height of observation point (receiver position) x=15m, current of the radiating Hertzian dipole I=1A, length of the Hertzian dipole 2h=0.1m (much smaller than the wavelength $\lambda$=c/f in both cases), relative dielectric constant of ground $\varepsilon_r$=20 and ground conductivity $\sigma$=0.01 S/m. Finally, note that the relation between current I and dipole moment p is given by : I(2h)=i$\omega$p , where $\omega$=2$\pi$f and i is the unit imaginary number.

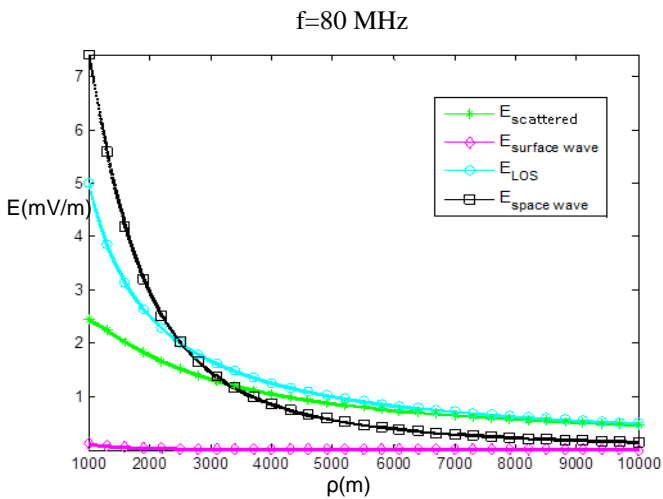

Fig. 3. Electric fields at observation point as a function of horizontal distance $\rho$ between transmitting Hertzain dipole and observation point, for frequency f=80 MHz.. Here the various components of received electric field are shown as following : Line – of – Sight (LOS) field (circle), field scattered from ground (asterisk), 'space wave' (square) and 'surface wave' (diamond). Note that in this case Norton's 'surface wave' is rather negligible as compared to the corresponding 'space wave' [10].

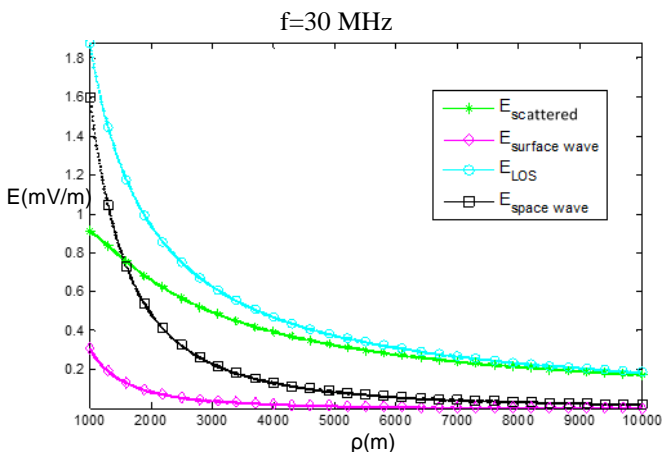

Fig. 4. Similarly with Fig. 3 above, except that here the frequency of radiating Hertzian dipole is now equal to 30 MHz (lower frequency). In this case of lower frequency the 'surface wave' can not be considered negligible, as compared to the 'space wave' [10].

## VI. CONCLUSIONS – FUTURE RESEARCH

In this paper we formulated the radiation problem from a vertical short (Hertzian) dipole above flat and lossy ground in the spectral domain, which resulted in an easy to manipulate integral expression for the received EM field above or below the ground. Subsequently, by applying the Stationary Phase Method (SPM) in the high frequency regime, the classical solution for the 'space wave' was rederived in a new fashion, thus showing that this is the dominant solution in this high frequency regime. Finally, numerical results in this high frequency limit were obtained, and they were compared to Norton's results for 'space' and 'surface' waves [3], [10].

Corresponding research in the near future by our research group will concentrate to the calculation of the received EM field below the ground at the high frequency regime (by using again the SPM method). Furthermore, and probably most important, to the calculation of the received EM field, above or below the ground, *for any frequency of the radiating dipole*, in an exact and analytical manner [11] or in a numerical way (i.e. through the use of numerical integration techniques). Moreover, we intend also to investigate the formulation of the same radiation problem in spectral domain, but now in the case of a *horizontal* radiating Hertzian dipole above flat and lossy ground. Finally, further investigations will be performed in the case of rough (and not flat) ground, in the case of curvature of the earth's surface for large distance communication applications etc.